\documentclass[useAMS,usenatbib]{mn2e}
\usepackage{graphicx}
\usepackage{txfonts}
\title[Accreting magnetars]{Accreting magnetars: a new type of high-mass 
X-ray binaries?}
\author[P. Reig et al.]{P. Reig$^{1,2}$\thanks{E-mail: pau@physics.uoc.gr},
    J.M Torrej\'on$^{3}$, P. Blay$^{4}$	\\
$^{1}$ IESL, Foundation for Research and Technology-Hellas, GR-71110 Heraklion,
Crete, Greece\\
$^{2}$ Institute of Theoretical \& Computational 
Physics, University of Crete, PO Box 2208, GR-710 03, Heraklion, Crete, Greece \\
$^{3}$ Instituto de F\'{\i}sica Aplicada a las Ciencias y las Tecnolog\'{\i}as, 
Universidad de Alicante, E 03080 Alicante, Spain  \\
$^{4}$ IPL, Universidad de Valencia, PO BOX 22085, 46071 Valencia, Spain}

\newcommand{\src}  {4U\,2206+54}
\newcommand{\ha}  {H$\alpha$}

\def\simless{\mathbin{\lower 3pt\hbox
     {$\rlap{\raise 5pt\hbox{$\char'074$}}\mathchar"7218$}}}   %< or of order
\def\simmore{\mathbin{\lower 3pt\hbox
     {$\rlap{\raise 5pt\hbox{$\char'076$}}\mathchar"7218$}}}   %> or of order

\def\msun{~{\rm M}_\odot}
\def\rsun{~{\rm R}_\odot}

\begin{document}

\date{Accepted ??. Received ??; in original form ??}

\pagerange{\pageref{firstpage}--\pageref{lastpage}} \pubyear{2012}

\maketitle

\label{firstpage}

\begin{abstract}

The discovery of very slow pulsations ($P_{\rm spin}=5560s$) has solved the
long-standing question of the nature of the compact object in the high-mass
X-ray binary 4U\,2206+54  but has posed new ones. According to spin
evolutionary models in close binary systems, such slow pulsations require a
neutron star magnetic field strength larger that the quantum critical value
of $4.4 \times 10^{13}$ G, suggesting the presence of a magnetar. We
present the first {\it XMM-Newton} observations of 4U\,2206+54 and
investigate its spin evolution. We find that the observed spin-down rate
agrees with the magnetar scenario. We analyse ISGRI/{\it INTEGRAL}
observations  of \src\ to search for the previously suggested cyclotron
resonance scattering feature at $\sim$30 keV. We do not find a clear
indication of the presence of the line, although certain spectra display
shallow dips, not always at 30 keV. The association of these dips with
a cyclotron line is very dubious because of its apparent transient nature.
We also investigate the energy spectrum of 4U\,2206+54 in the energy range
0.3-10 keV with unprecedented detail and report for the first time the
detection of very weak 6.5 keV fluorescence iron lines.  The photoelectric
absorption is consistent with the interstellar value, indicating  very
small amount of local matter, which would explain the weakness of the
florescence lines. The lack of matter locally to the source may be the
consequence of the relatively large orbital separation of the two
components of the binary. The wind would be too tenuous in the vicinity of
the neutron star. 

\end{abstract}


\section{Introduction}

The X-ray source \src\ is a peculiar member of the high-mass X-ray binary
systems (HMXB). In this type of X-ray binaries, the normal stellar
component is a massive star, usually an O or B star, while the X-ray
emitting component is generally a neutron star. Mass transfer in HMXBs may
occur via the strong stellar wind of the optical companion as in systems
containing supergiant stars, the so-called supergiant X-ray
binaries\footnote{Accretion via an accretion disk is believed to occur in
only three SGXBs due to Roche lobe overflow.} (SGXB) or via the disc-like
envelope of photospheric material around the equator of main-sequence
companions.  In this latter case, the system is called a Be/X-ray binary
(BeXB). Wind-fed systems are normally persistent X-ray sources with X-ray
luminosities of the order of $10^{35}-10^{36}$ erg s$^{-1}$, while most
BeXB are transient sources with episodes of violent eruption in X-rays that
may reach $10^{38}$ erg s$^{-1}$. The optical/X-ray properties of \src\ do
not fit in any of these two categories of HMXBs. 

It is classified as a HMXB because it contains a neutron star orbiting a
O-type companion \citep{negueruela01}. However,  \src\ does not belong to
the supergiant subgroup because the optical companion is a main-sequence
star \citep{blay06}. The absence of any obvious long-term trends in
the evolution of the \ha\ line spectral parameters and the lack of any
correlation between the \ha\ line equivalent width and infrared magnitudes
and colours  differ from what it is expected in a typical BeXB. 
Optical  and infrared studies indicate that the optical companion in \src\
is an O9.5V star located at a distance of 2.6 kpc \citep{blay06} with
a higher than normal He abundance. The X-ray emission is the result of
accretion from an abnormally slow wind \citep{ribo06}.  

For a long time the nature of the compact companion was debated. Recently,
\citet{reig09} discovered a 5560-s modulation in the X-ray light curve that
was interpreted as the spin period of the neutron star.  The discovery of
very slow pulsations has solved the controversy on the nature of the
compact object in \src,  but has prompted new questions. According to 
spin evolutionary models in close binary systems, such slow pulsations
require magnetic field strengths of the order of $\simmore 10^{14}$ G, that
is, they require the system to harbour a magnetar \citep{finger10,ikhsanov10}. 

The magnetic field strength can be estimated considering the neutron star
spin evolution. \citet{finger10}  derived an average rate of the spin
frequency change of  $\dot{\nu} = (-1.7\pm0.3) \times 10^{-14}$ Hz
s$^{-1}$. Based on the slow pulsation period and the spin down rate, they
show that  the 5560 s pulsations in the X-ray flux of \src\ can be
explained "provided the neutron star in this system is a magnetar whose
surface field at the present epoch exceeds $10^{14}$ G".

However, these results are at odds with the tentative detection of a cyclotron
resonance scattering feature (CRSF).  Although the statistical significance
of this feature is only marginal in any single one measurement, it has been
reported by three different observatories: {\it RXTE} \citep{torrejon04},
{\it BeppoSAX} \citep{masetti04}, and {\it INTEGRAL} \citep{blay05,wang09}
at different epochs. However, no sign of this feature has been detected in
other observations with the same instruments on board {\it RXTE}
\citep{reig09} and {\it INTEGRAL} \citep{wang10}. If real, the detection of
a CRSF at $\sim30$ keV would imply a magnetic field strength ($B/10^{12}
{\rm G}\approx E_{\rm cycl}/11.6 {\rm keV}$) of $3.3\times 10^{12}$ G,
typical of accretion-powered pulsars.

%There is growing evidence supporting that the presence of a soft thermal
%component is an ubiquitous property in all accreting pulsars
%\citep{hickox04}. In BeXB at low luminosity ($L_X \simless 10^{35}$ erg
%s$^{-1}$) this soft component is believed to represent emission from the
%polar caps \citep{palombara10}.

In this work, we analyse the first {\it XMM-Newton} observations of \src.
We confirm the existence of the 5560-s X-ray pulsations, derive a new value
of the spin-down rate and perform a detailed spectral study of the soft
emission with unprecedented sensitivity. We also analyse ISGRI/{\it
INTEGRAL} observation to search for the presence of the cyclotron line.

\section{Observations and data analysis}

%We have performed a timing and spectral analyses of the HMXB \src\ using 
%{\it XMM-Newton} and INTEGRAL data. The {\it XMM-Newton} observations
%represent the first of this source. 

\subsection{XMM-Newton observations}

\src\ was observed by XMM-Newton on 2011 February 6 during revolution 2044.
The observation (ObsID 0650640101) started at 03:39 hr UT and lasted for
$\sim$75 ks. The observation details for the instruments on-board
XMM-Newton are presented in Table~\ref{obs}. The XMM-Newton Observatory
\citep{jansen01} includes three 1500 cm$^2$ X-ray telescopes each with an
European Photon Imaging Camera (EPIC) at the focus. Two of the EPIC imaging
spectrometers use MOS CCDs \citep{turner01} and one uses PN CCDs
\citep{struder01}. Reflection Grating Spectrometers
\citep[RGS][]{denherder01} are located behind two of the telescopes while
the 30-cm optical monitor (OM) instrument has its own optical/UV telescope
\citep{mason01}. All instruments were in operation during the observation.
Data were reduced using the XMM-Newton Science Analysis System (SAS version
11.0).

We first reprocessed Observation Data Files (ODFs) to obtain calibrated and
concatenated event lists.  The EPIC-PN analysis was performed on this ODF
event file. We restricted the useful PN events to those with a pattern in
the range 0 to 4 (single and doubles) and complying with the more strict
selection criterion FLAG=0, which omits parts of the detector area like
border pixels and columns with higher offset.

Although the observation were made in "partial window" mode, the brightness
of the source was so high that the EPIC instruments were affected by pile-up.
Because the PN camera has a faster read-out time, it is less affected by
this effect. Consequently, we used PN data in our timing and broad-band
spectral analysis.  PN spectra and light curves were generated by excluding
the core of the PSF. The same extraction region was used for both light
curves and energy spectra. We extracted events from an annulus region with
inner and outer radii 10 and 60 arcsec, respectively. Background files were
extracted from a circular source-free region with a radius 30 arcsec
centered approximately at the same distance as the readout node (RAWY) as
the source region. To achieve the best energy resolution and minimise
further the effect of pile up the final PN spectrum was created from single
events only. The SAS tasks {\em rmfgen} and {\em arfgen} were used to
generate source-specific response matrix files and auxiliary response files
for spectral analysis. We rebinned the EPIC-PN spectrum to oversample the
full-width at half maximum (FWHM) of the energy resolution by a factor 3,
and to have a minimum of 25 counts per bin to allow the use of the
$\chi^{2}$ statistic.  

The SAS task {\em rgsproc} was used to produce calibrated RGS event lists,
spectra, and response matrices. The individual spectra of the same spectral
order obtained from each one of the two RGS instruments were combined ({\em
rgscombine}) to produce the final RGS spectrum.

%------------------------------------------------------------------------------
\begin{table}
\begin{center}
\caption{XMM-Newton observation log.}
\label{obs}
\begin{tabular}{cccc}
\hline \hline \noalign{\smallskip}  
Instrument	&Mode	&Filter	&Exposure (ks)	\\
\hline \noalign{\smallskip}
EPIC-PN	&large window	&medium	&74.8	\\
MOS1	&small window	&medium	&72.4	\\
MOS2	&small window	&medium	&72.4	\\
RGS	&standard	&--	&76.4	\\
OM	&image		&grism1	&71.7	\\
\hline \hline \noalign{\smallskip}
%\multicolumn{5}{l}{$a$:}\\
\end{tabular}
\end{center}
\end{table}
%------------------------------------------------------------------------------

\subsection{INTEGRAL observations}

INTEGRAL (INTERnational Gamma-RAy Laboratory) is an ESA space mission with
contributions from NASA and Russia dedicated to the observation of the hard
X-rays/soft-$\gamma$-ray Universe \citep{winkler03}. The two main
instruments on board INTEGRAL are IBIS (Imager on Board Integral
Spacecraft) which operates in the 15 keV--10 MeV energy range 
\citep{ubertini03} and SPI (SPectometer on board Integral) which operated
in the 20 keV--8 MeV energy range \citep{vedrenne03}. IBIS, in turn, is
composed of two detector layers, ISGRI (Integral Spacecraft Gamma-Ray
Imager) operational in the 15-1000 keV range, and PCSiT (Pixellated CSi
Telescope), working in the 0.175-10 MeV range 
\citep{lebrun03,labanti03}. At lower energies, the
payload of {\it INTEGRAL} consists of the X-ray monitor JEM-X, sensitive to
X-rays in the 3-35 keV band \citep{lund03} and the optical monitor OMC that
covers the visual range 5000-6000 \AA\ \citep{mas03}.

INTEGRAL observations normally consist of a series of pointings called Science
Windows (SCW), 2 to 4 ks each. Since the launch of INTEGRAL in October 2002,
\src\ has repeatedly appeared in the field of view of its instruments. However,
the number of observations with a significant detection of the source (those
with a detection level above 7$\sigma$) is relatively small. We have analysed all
publicly available ISGRI observations from the beginning of the mission up to
revolution 966 (September 2010). We selected the SCW where the source was within
10 degrees from the centre of the pointing and the detection level was above 7.
This selection resulted in 37 SCWs.  For each SCW an ISGRI spectrum was
obtained. These spectra were combined together to produce an average ISGRI
spectrum. The Integral Science Data Center (ISDC) software OSA (Offline
Scientific Analysis) version 9.0 was used in the data processing. 

%-------------------------------fig 1-----------------------------------------------
\begin{figure}
\resizebox{\hsize}{!}{\includegraphics{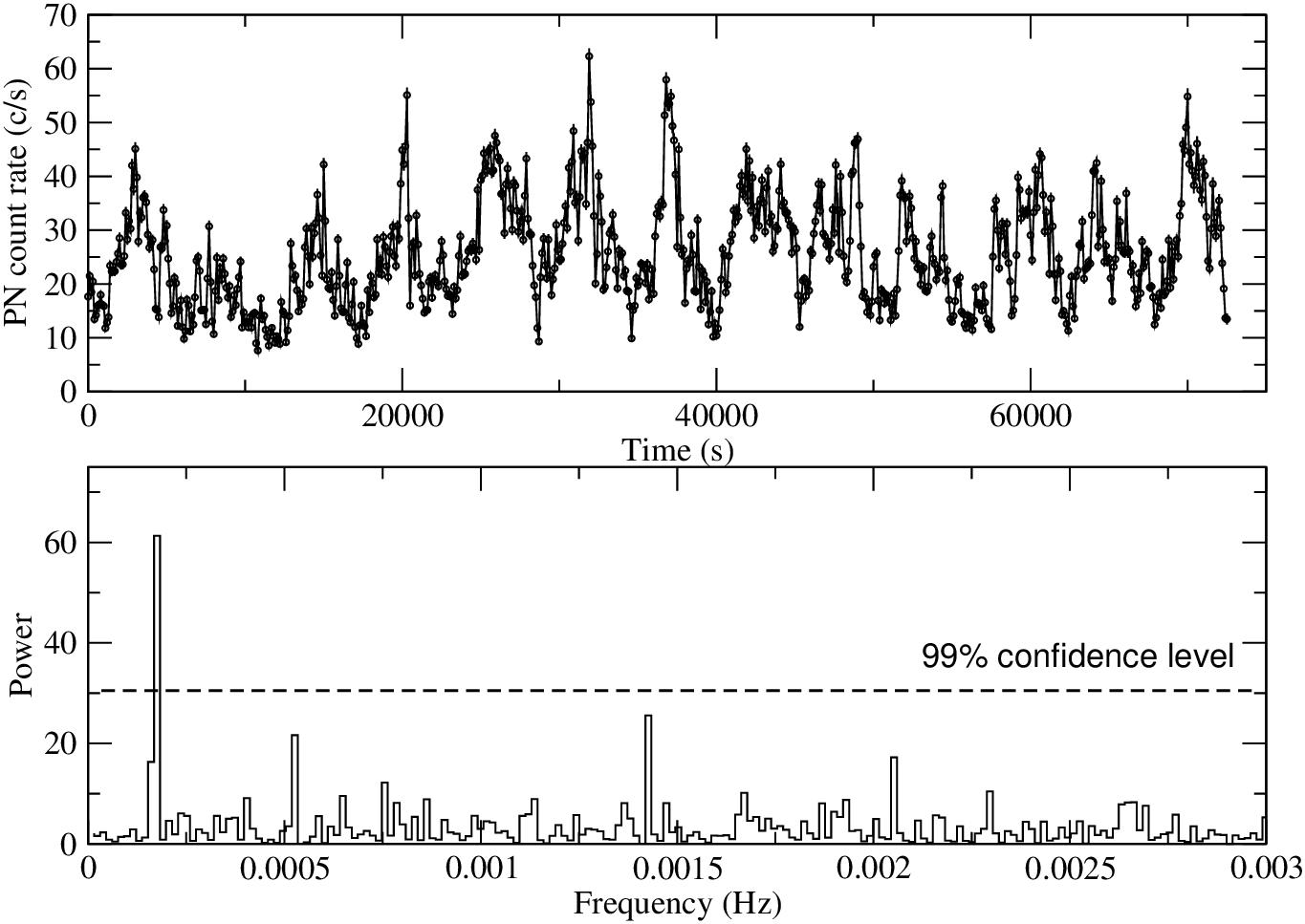} } 
\caption[]{{\em Upper panel}: PN light curve \src. Time zero is MJD
55598.17 and bin size is 100 seconds. {\em Bottom panel}: FFT using a
10-s binned light curve. A peak at $\sim$ 0.17 mHz, representing the 5590-s
pulsations, is clearly seen. }
\label{pn_lc_psd}
\end{figure}
%------------------------------------------------------------------------------

\section{The spin period evolution}

Prior to the determination of the spin period, we converted the arrival
times to the solar system barycenter. Then we generate a light curve in the
0.2-12 keV energy range with a bin size of 10 seconds. The last $\sim$2300
seconds of the light curve were affected by high background. After removing
this part the total duration of the light curve was 72.5 ks.

Figure~\ref{pn_lc_psd} shows the 0.2-12 keV PN light curve rebinned to 100-s
bins and the Fast Fourier transform (FFT) obtained from the 10-s binned
light curve. A coherent periodicity is visible in the light curve and
confirmed by the FFT. The maximum power is distributed around
$\sim$0.0001775 Hz, corresponding to a period of $\sim$5635 s. The power
spectrum of \src\ is characterised by strong red noise at low frequencies,
which is well described by a power-law model with a slope of 1.5--1.7
\citep{reig09}. To assess the statistical significance of this peak and
give a confidence detection limit (dashed line in Fig.~\ref{pn_lc_psd}), we
first corrected for the presence of this type of noise by fitting a
power-law function to the continuum. The power spectrum shown in
Fig.~\ref{pn_lc_psd} was created  by multiplying the original power
spectrum by 2 and dividing by the power-law model.

The small number of pulse cycles covered by the light curve ($\sim$13)
prevented us from using the pulse phase connection technique \citep[see
e.g.][and references therein]{staubert09} to derive an accurate value of the 
spin period because the pulse profiles would not have sufficient statistics. 
Nevertheless, we can use other less accurate
techniques to refine the value of the spin period. We used the CLEAN
\citep{roberts87}, Lomb-Scargle \citep{scargle82} and Phase Dispersion
Minimization \citep[PDM][]{stellingwerf78} algorithms. All these algorithms
are implemented in the program PERIOD (version 5.0), distributed with the
Starlink Software Collection. Table~\ref{algo} gives the results of the
timing analysis. Another technique
that achieves smaller errors is epoch folding \citep{leahy83,leahy87}. We
used the task {\it efsearch} of the XRONOS package and equation (6a) in
\citet{leahy87} to estimate the error. We obtained $P_{\rm spin}=5591\pm3$
s. One more determination of the spin period was done with the
software package PERIOD04 \citep{lenz05}, which uses a discrete
Fourier transform (DFT) algorithm and allows the estimation of
uncertainties through Monte Carlo simulations. This method gave $5600\pm5$
s, where the error was estimated from the distribution of frequencies after
running 1000 simulations. The mean spin period obtained as the average of
the six methods is $5585\pm30$ s, where here the error is the standard
deviation of the measurements. A weighted mean gives $5593\pm3$ s.
However, it should be noted that these calculations apply in the case that,
in the absence of pulsations, the only noise present in the data is that
due to counting statistics, i.e, white noise. Because \src\ shows strong
red noise in its power spectrum \citep{reig09}, \citet{finger10} used a
more realistic approach and assumed that the natural log of the rates is
(as opposed to just the rates) Gaussian distributed. Then they used the
{\em Cash} statistics and compared it to the results obtained with the
Lomb-Scargle method. They found that the pulse periods estimated from the
peak of the Lomb-Scargle periodogram were shifted by 1--20 s (depending on
the initial light curve) with respect to the maximum likelihood fit
(minimisation of the {\em Cash} statistic). Taking into account these
statistical considerations we adopted a final value of $P_{\rm}=5593\pm10$
s.

\begin{figure}
\resizebox{\hsize}{!}{\includegraphics{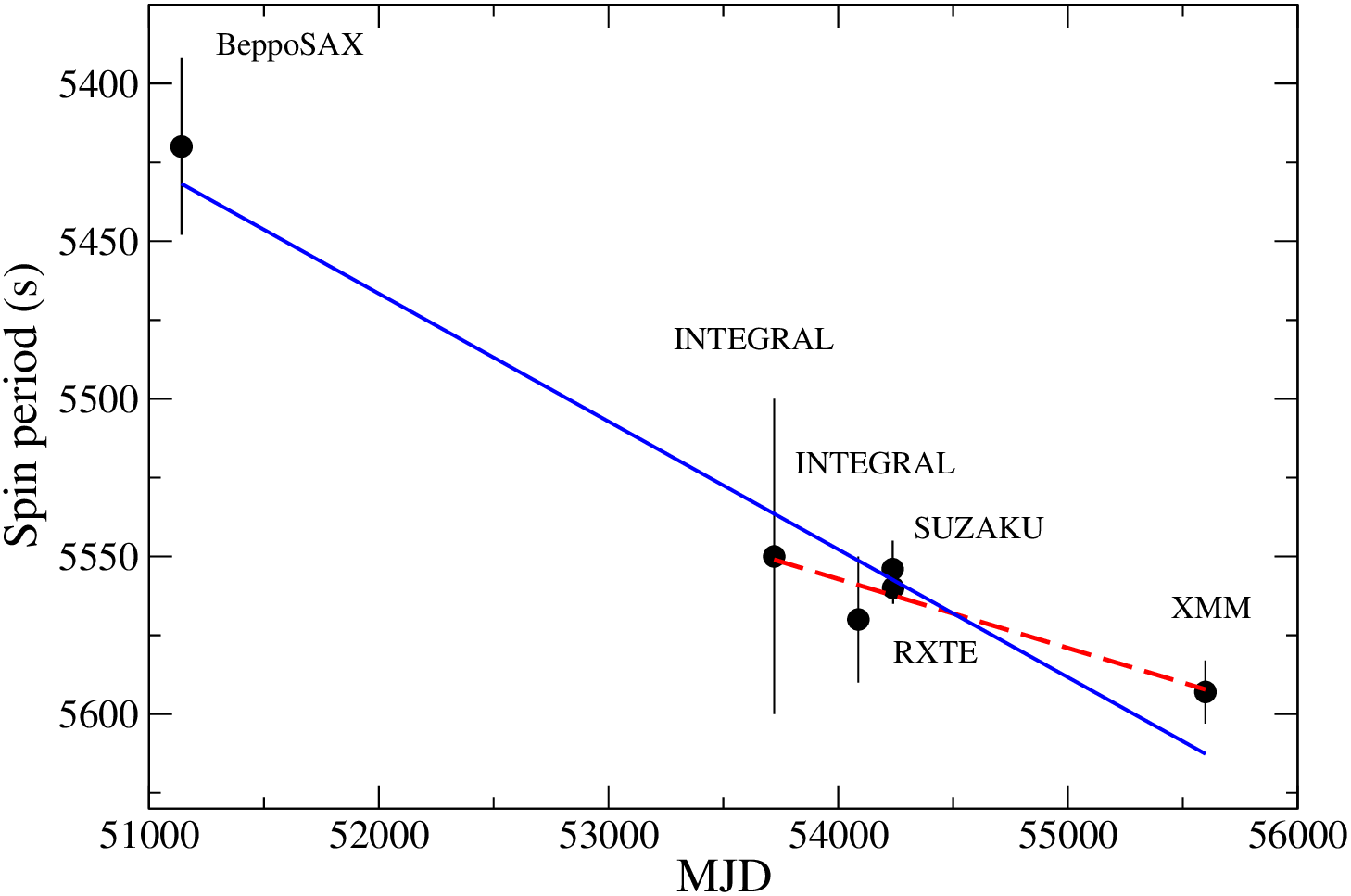} } 
\caption[]{Spin-down evolution. The lines represent the best linear
fit to all the data (solid line) and after removing the oldest data point 
(dashed line). }
\label{spinevol}
\end{figure}

Figure~\ref{spinevol} shows the spin-down evolution of \src. The data
points come from \citet[][BeppoSAX \& Suzaku]{finger10}, \citet[][INTEGRAL
\& RXTE]{reig09} and \citet[][INTEGRAL]{wang10}. Using only the
{\it BeppoSAX} and {\it Suzaku} measurements of the spin period,
\citet{finger10} obtained a spin-down rate of $\dot{\nu}=(-1.7\pm0.3)
\times 10^{-14}$ Hz s$^{-1}$. Figure~\ref{spinevol} shows that this trend
continues up to present. A linear fit to all the data points gives
$\dot{\nu}=(-1.5\pm0.2) \times 10^{-14}$ Hz s$^{-1}$ with a correlation
coefficient of 0.97 (solid line). 
%---------------------------------fig2---------------------------------------------
%------------------------------------------------------------------------------

%------------------------------------------------------------------------------
\begin{table}
\begin{center}
\caption{Spin period determination.}
\label{algo}
\begin{tabular}{lc}
\hline \hline \noalign{\smallskip}  
Method		&$P_{\rm spin}$ (s)	\\
\hline \noalign{\smallskip}
%FFT		&5690$\pm$111	\\
CLEAN		&5580$\pm$107	\\
Lomb-Scargle	&5580$\pm$107	\\
PDM		&5532$\pm$105	\\
\hline
DFT		&5600$\pm$5		\\
\hline
Epoch folding	&5591$\pm$3	\\
\hline
Mean		&5586$\pm$30	\\
Weighted mean	&5593$\pm$3	\\
\hline
Final adopted value	&5593$\pm$10	\\
\hline \hline \noalign{\smallskip}
%\multicolumn{5}{l}{$a$:}\\
\end{tabular}
\end{center}
\end{table}
%------------------------------------------------------------------------------

%------------------------------------------------------------------------------
\begin{table}
\begin{center}
\caption{Best-fit spectral parameters to the EPIC-PN spectrum. 
The luminosity and blackbody emitting radius were computed assuming a 
distance of $2.6\pm0.3$ kpc \citep{blay06}. Quoted errors are at 
90\% confidence level for a single parameter. ABS: photoelectric absorption, 
BB: blackbody, 
BMC: Bulk Motion Comptonization, PL: power law. }
\label{specfit}
\begin{tabular}{lc}
\hline \hline \noalign{\smallskip}
\multicolumn{2}{c}{ABS*(BB+BMC)}	 \\
\hline \noalign{\smallskip}
$N_H$ ($\times10^{21}$ cm$^{-2}$)	&2.23$\pm$0.02	\\
$kT_{\rm bb}$ (keV)			&0.51$\pm$0.02   \\
$R_{\rm bb}$ (km)			&1.3$\pm$0.2	\\
$kT_{\rm col}$ (keV)			&1.47$^{+0.07}_{-0.04}$	\\
$\alpha$				&1.37$^{+0.05}_{-0.4}$	\\
$L_{bb}$ ($\times10^{35}$ erg s$^{-1}$)	&0.13		\\
$L_{BMC}$ ($\times10^{35}$ erg s$^{-1}$)&2.7		\\
$\chi^2$/dof				&0.98/197	\\
\hline
\multicolumn{2}{c}{ABS*(BB+PL)}	\\
\hline
$N_H$ ($\times10^{21}$ cm$^{-2}$)	&3.72$\pm$0.08	\\
$kT_{\rm bb}$ (keV)			&1.63$\pm$0.03	 \\
$R_{\rm bb}$ (km)			&0.37$\pm$0.04	 \\
$\Gamma$				&0.94$\pm$0.03	\\
$PL_{\rm norm}^a$			&1.11$\pm$0.04	 \\
$L_{bb}$ ($\times10^{35}$ erg s$^{-1}$)	&1.17		\\
$\chi^2$/dof				&1.25/200	  \\
\hline
\multicolumn{2}{c}{ABS*(BB+BB)}	\\
\hline
$N_H$ ($\times10^{21}$ cm$^{-2}$)	&2.00$\pm$0.05	\\
$kT_{\rm bb1}$ (keV)			&0.69$\pm$0.02 	\\
$R_{\rm bb1}$ (km)   			&1.1$\pm$0.1    \\
$kT_{bb2}$ (keV) 			&2.11$\pm$0.02 \\
$R_{\rm bb2}$ (km)   			&0.34$\pm$0.04     \\
$\chi^2$/dof				&1.45/200	  \\
\hline
$L_{\rm 0.3-2keV}$ ($\times10^{35}$ erg s$^{-1}$)	&0.16   \\           
$L_{\rm 2-10keV}$ ($\times10^{35}$ erg s$^{-1}$)	&2.21	\\
\hline \hline \noalign{\smallskip}
%\multicolumn{3}{l}{ABS: photoelectric absorption, BB: blackbody, BMC: Bulk
%Motion Comptonization, PL: power law}\\
%\multicolumn{4}{l}{$a$:$\times10^{35}$}\\
\multicolumn{2}{l}{$a$:$\times10^{-2}$ ph keV$^{-1}$cm$^{-2}$s${^1}$ at 1 keV}\\
%\multicolumn{4}{l}{$c$:$\times10^{35}$}\\
\end{tabular}
\end{center}
\end{table}
%------------------------------------------------------------------------------

%---------------------------------fig3---------------------------------------------
\begin{figure}
\resizebox{\hsize}{!}{\includegraphics{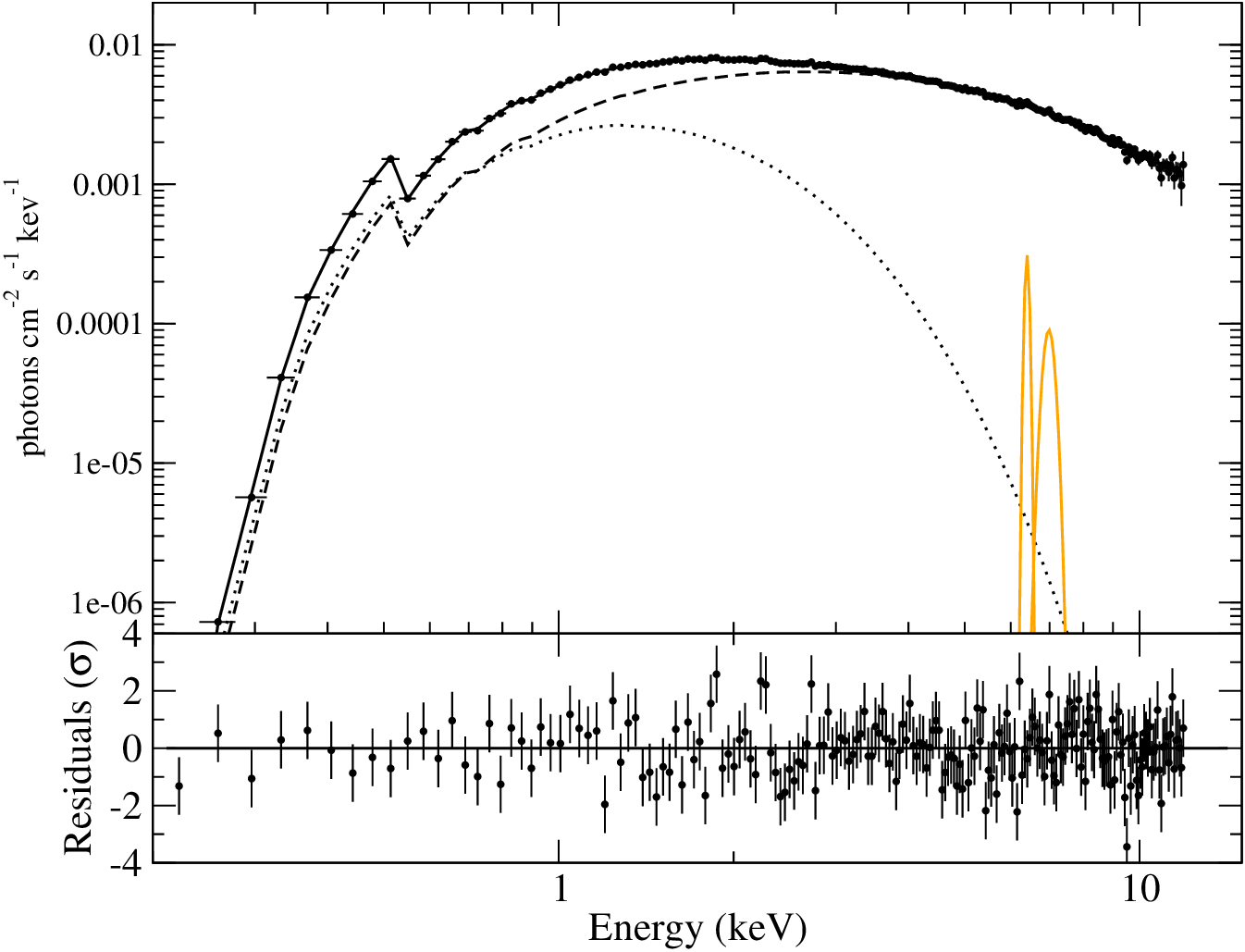} } 
\caption[]{EPIC-PN energy spectrum. The BMC (dashed line) dominates the 
description of the continuum spectrum at high energies while a blackbody 
(dotted line) has been added to correct the excess at low energies.}
\label{pn-spec_bmc}
\end{figure}
%------------------------------------------------------------------------------

\section{The XMM-Newton spectrum}

The XMM observations allows us to investigate the low-energy part of the
spectrum in \src\ with unprecedented sensitivity.

\subsection{The EPIC-PN spectrum}

\label{softx}

Figure~\ref{pn-spec_bmc} shows the EPIC-PN spectrum of \src, covering the
energy range 0.2-12 keV. This spectrum can be well represented by
two-component models, affected by photoelectric absorption. The first component
is a blackbody, which accounts for the softer part of the spectrum, while
the second component dominates above $\sim$3 keV. Good fits to the EPIC
spectral continuum are obtained when the second component is represented by
a power law, by Bulk Motion Comptonization \citep[BMC,][]{shrader98} or
even by another blackbody. The best-fit parameters of these two combination
of models are presented in Table~\ref{specfit}. 

The fit with the reduced $\chi^2$ closest to 1 is
obtained with the combination that includes the BMC. In this model, the
power law spectral index $\alpha$ yields a value larger than 1 which means
that the Comptonization efficiency is high. Consequently, the
illumination factor $f$, defined as the number of photons multiply
scattered with respect to the number of injected photons by the thermal
component, is $>>1$, that is to say, the spectrum appears completely
comptonized. The size of the seed soft photons production site can be
estimated from the luminosity of the soft photons source
$L_{0}=\frac{L}{\ln(1/x_{0})}$ where $L$ is the resulting comptonized
luminosity and $x_{0}=2.8kT_{col}/kT_{e}$ \citep{titarchuk97}. Using the
values in Table~\ref{specfit}, and assuming a comptonizing cloud of
$kT_{e}\sim 50$ keV \citep[as obtained in previous works,
e.g.][]{torrejon04} we obtain $L_{0}=1.1\times 10^{35}$ ergs s$^{-1}$. If the
soft photon source is radiating like a blackbody of area $\pi R_{\rm
W}^{2}$ then we have

\begin{displaymath}
R_{\rm W} = 0.6 \sqrt{L_{0}^{34}}(kT_{\rm col}/1~\rm keV)^{-2}~[km]
\end{displaymath}

\noindent which yields $R_{\rm W}\approx 1$ km. This is consistent with the
size of a hot spot over the neutron star surface. On the other hand, if we
do not assume any temperature for the comptonizing cloud and use instead
the total luminosity for the BMC component, the resulting radius is 1.6 km,
still very small and compatible with a hot spot. 

Alternatively, the size of the emitting region can be estimated from  the
blackbody component. The luminosity of the soft excess turns out to be
$L_{bb}=1.3\times 10^{34}$ ergs s$^{-1}$ which, for the temperature given
in Table~\ref{specfit}, yields a circular radius\footnote{If the
interpretation that the blackbody component represents emission from hot
spots is correct, then it seems more natural to assume a circular area as
the emitting region. However, the values found in the literature normally
assume a spherical emitting region. For this reason, the radii in
Table~\ref{specfit} correspond to a spherical emitting area.  Note that if
the soft photons source is taken to radiate as a blackbody {\it spherical}
surface (i.e. $4\pi R_W^2$), the corresponding radius would be {\it half}
that obtained from a circular region. } of $R_{bb}\approx 2.4$ km. This is
again too small to be produced by any material surrounding the neutron
star. This would be consistent with a reprocessing region (ring)
surrounding the hot spot, the accretion mound or the top of the shock front
at the impact point and excludes the innermost parts of a hypothetical
accretion disk.

Given the limited band pass of XMM-EPIC cameras, one might be worried about
what fraction of the spectrum is being really accounted for by 
Comptonization. As can be seen in Fig.~\ref{pn-spec_bmc}, the BMC
component is fitting the high energy part of the spectrum while the
blackbody is contributing, mainly, at energies below 3 keV. Nonetheless, it
is perhaps convenient to give the fitting in terms of a phenomenological
model composed by the addition of a power law and a blackbody modified at
low energies by photoelectric absorption. These parameters are given in the
middle part of Table \ref{specfit}. As can be seen, the fit is
acceptable, albeit with a higher value of the reduced $\chi^{2}$. The most
noticeable change is the increase in temperature of the blackbody component
which is now contributing at higher energies (peak at 4 keV). The radius of
the soft photon source would be $R_{bb}\simeq 0.7$ km (0.37 km if an spherical
emitting region is considered),  smaller than, but of the order of, that
found under the Comptonization interpretation.

The radii obtained from the two-blackbody model are also consistent with
emission from the polar caps (Table~\ref{specfit}), although in this case
the model shows the largest residuals.

The error in $R_{\rm bb}$ was obtained by propagating the errors from the
best-fit blackbody normalisation, $N$, ($R_{\rm bb}= \sqrt{N} D_{10}$)) as
given by the $\chi^2$ fit and the error on the distance, which was
estimated to be 0.3 kpc \citep{blay06}. $D_{10}$ is the distance to
the source normalised to 10 kpc.

A high signal-to-noise spectrum at low energies is also particularly
interesting because it allows the determination of the hydrogen column
density, $N_H$. The energy range 0.1-2.0 keV is the most suitable  band for
the derivation of the $N_H$ since cold gas mainly absorbs X-ray photons
with energies lower than 2--3 keV.  We find $N_H=(0.36\pm0.01)\times
10^{22}$, which is the most accurate determination of the absorption to the
source measured so far. The $N_H$ allows us to derive the optical
extinction to the system. Using the calibration by \citet{guver09}, we
obtain $A_V=1.63\pm0.08$ mag or, assuming a standard reddening law,
$E(B-V)=A_V/3.1= 0.53$ mag, in excellent agreement with the value of
0.54$\pm$0.05 mag derived from optical observations \citep{blay06}. This
value agrees very well with the reddening derived using diffuse
interstellar bands, hence it is consistent with Galactic absorption along
the line of sight to \src. We conclude that there is very little matter
locally.

%-----------------------------------------------------------------------
\begin{table}
\caption{RGS model parameters.}            
\label{tab:rgs}      
\centering                          
\begin{tabular}{ l c c  }        
\hline\hline   
            \noalign{\smallskip}          
 Parameter & ABS*(BB)  & ABS*(BB+PO) \\
            \noalign{\smallskip}
\hline    
            \noalign{\smallskip}                  
$N_{H}$($\times 10^{22}$ cm$^{-2}$)   & 0.10$\pm$0.02 & 0.16$\pm$0.04 \\
$kT_{bb}$ (keV)   &2.10$\pm$0.30  & 1.95$\pm$0.03 \\
Norm$_{bb}$   &   0.0098$\pm$0.0027  &    0.0082$\pm$0.0004\\
$\chi^{2}_{\nu}$ (dof) & 1.19 (702) & 1.19 (701)\\
            \noalign{\smallskip}
\hline                                  
\end{tabular}
\begin{list}{}{}
\item Power-law photon index fixed at $\Gamma=0.97$
\end{list}

\end{table}
%-----------------------------------------------------------------------

%-----------------------------------fig4------------------------------------
\begin{figure}
   \centering
   \includegraphics[width=0.75\columnwidth, angle=-90]{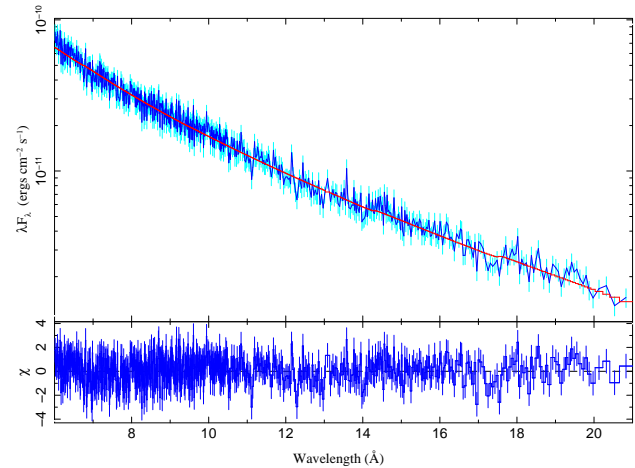}
  \caption{RGS spectrum. }
    \label{rgs}
\end{figure}  
%-----------------------------------------------------------------------

\subsection{RGS spectrum}

The low value of the reddening is further supported by the analysis of the
RGS data. In Fig. \ref{rgs} we show the RGS  spectrum of the source
in the wavelength range 6 to 20 \AA\ (= 0.6 to 2 keV). The data from RGS 1
and 2 have been added and the resulting spectrum rebinned so that each bin
has a minimum signal-to-noise ratio of 7. In order to detect the presence
of any possible lines, the modeling of the underlying continuum is
critical. Therefore we take advantage of the broader energy range of PN and
use the parameters deduced in the previous section as input. In this
wavelength range the main contributor is the blackbody. Consequently we 
tried a single absorbed blackbody as well as an absorbed power law plus
blackbody.  In the latter case, the power law parameters were frozen while
those of the blackbody were let to vary freely. The best-fit parameters are
given in Table~\ref{tab:rgs}. Two important conclusions can be drawn from
this analysis. First, no emission lines from an ionized plasma are
detected. This is in agreement with the PN analysis where the emission line
candidates from photoionized species, if present, are extremely weak (see
Sect.~\ref{lines}). Second, within the RGS interval the main indicators of
the photoelectric absorption are the Ne edge at 14.2 \AA\ and the Fe L edge
at 17.5 \AA.  As can be seen, the edges are extremely shallow. This supports
the low value of the photoelectric absorption obtained from the analysis of
PN data.

%-----------------------------------------------------------------------
\begin{table}
\caption{Emission line parameters.}            
\label{table:lines}      
\centering                          
\begin{tabular}{ l c c c c}        
\hline\hline   
            \noalign{\smallskip}          
Line &  E   &  $\sigma$ & Flux & EW \\
       &     &     & ($\times 10^{-5}$) &   \\
              &  (keV)    & (keV)    & ph s$^{-1}$ cm$^{-2}$   &  (eV)  \\
\hline    
            \noalign{\smallskip}                  
Fe K$\alpha$ & 6.46$\pm$ 0.05  & 0.14$\pm$ 0.06 & 8.9$\pm$3.0 & 25.0$\pm$8.4\\
Fe \textsc{xxvi} Ly $\alpha$ & 6.99$\pm$ 0.02 & 0.14 & 5.2$\pm$2.0 & 16.5$\pm$13.2 \\
Si \textsc{xiv} Ly $\alpha$ & 2.22$\pm$0.05 & 0.05 & 6.4$\pm$2.0 & 7.4$\pm$2.5\\
            \noalign{\smallskip}
\hline                                  
\end{tabular}
\begin{list}{}{}
\item Numbers without errors were fixed at the quoted values
\end{list}

\end{table}
%-----------------------------------------------------------------------

\subsection{Detection of emission lines}

\label{lines}

The X-ray spectrum of 4U2206+54 seems to be devoid of any strong emission
features. The high sensitivity of XMM, however, allowed us to detect,
for the first time, weak emission lines. The residuals are clearly
structured around the Fe complex region, at 6.5 keV and also around 2 keV.
The region around 6.5 keV is shown in Fig. 6. These features can be fitted
with two gaussians centred at 6.46$\pm 0.05$ keV and at 6.99$\pm$0.02 keV
respectively. Both lines show a width of $\sigma=0.14$ keV but only in the
first case the errors could be constrained ($\pm 0.06$) while we fixed the
width for the second one. The lines are significant at three sigma. When
the gaussians are added, the $\chi^{2}$ of the fit decreases from 251 for
202 degrees of freedom to 214 for 198 degrees of freedom. The energy of
these components is consistent with the Fe K$\alpha$ (6.4 keV) transition
from neutral Fe and Fe XXVI Ly$\alpha$ (6.96 keV) transition from ionized
H-like Fe. The parameters of these lines are shown in Table
\ref{table:lines}. As can be seen , even though the detection is firm, the
equivalent width is very small (25 eV and 16.5 eV, respectively).
Consistently with the extreme weakness of the Fe K$\alpha$ line, no Fe edge
at 7.14 keV is detected.

The residuals around the 2 keV region can be fitted with a narrow Gaussian at
2.22$\pm 0.05$ keV. The centroid energy is consistent with the Si XIV Ly$\alpha$
transition of H-like Si. The line appears very narrow and can not be
resolved by the instrument. We fixed the width to 0.05 keV. The
significance of this line is, however, low  (2$\sigma$).

%-------------------------------fig5 -----------------------------------------------
\begin{figure}
\resizebox{\hsize}{!}{\includegraphics{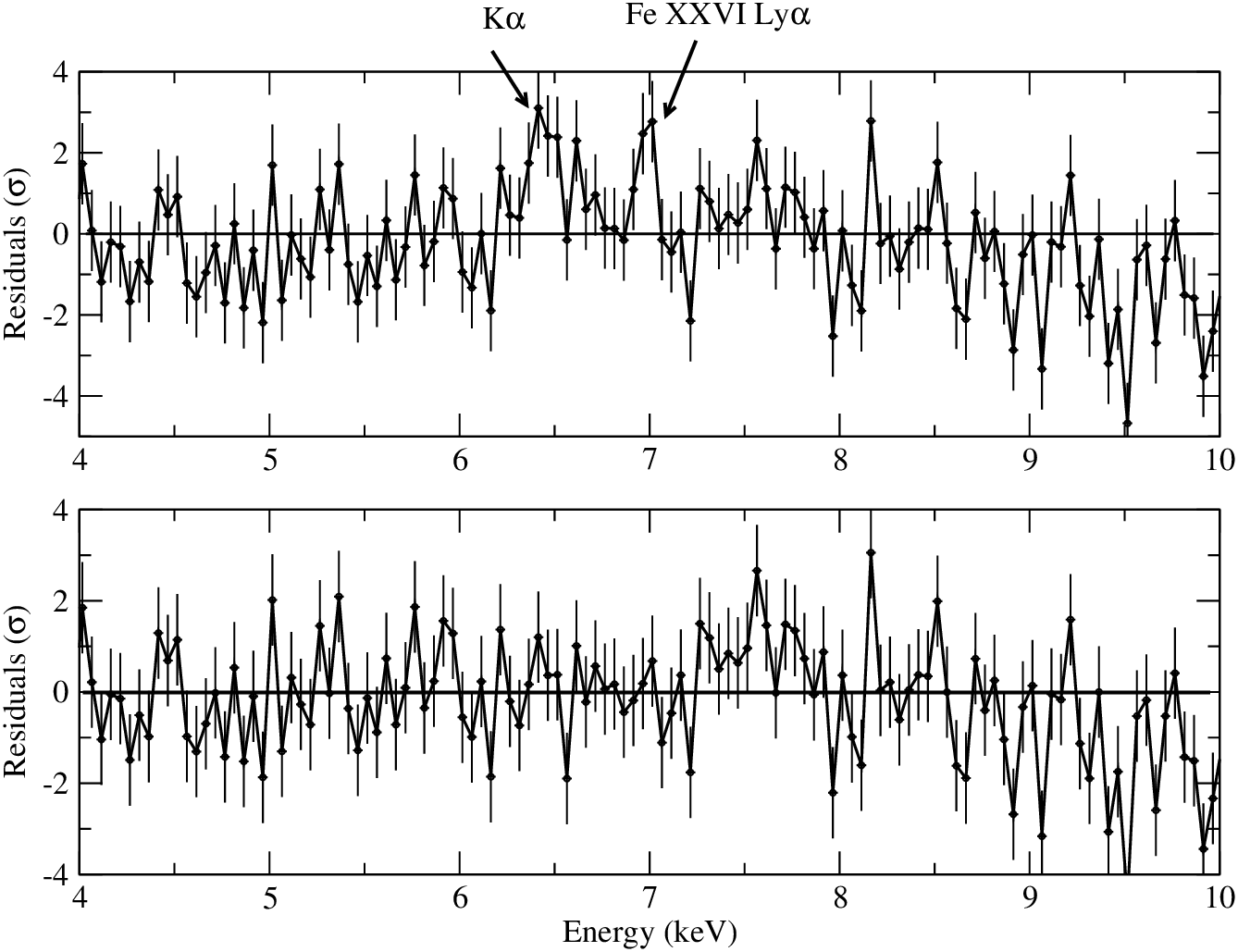} } 
\caption[]{Residuals in number of sigmas of the fit to EPIC-PN energy spectrum 
around the iron complex region. The lower and upper panels show the residuals 
with and without the inclusion of two Gaussian profiles, respectively }
\label{pn-iron}
\end{figure}
%------------------------------------------------------------------------------

\section{The ISGRI spectrum}

Previous works \citep{torrejon04,masetti04,blay05,wang09} have reported the
possible presence of a cyclotron resonant scattering feature (CRSF) at
around 30 keV. All the detections of this feature were tentative because
the statistical significance of any individual detection was $\simless
2\sigma$.  Since the gamma-ray imager on board {\it INTEGRAL} is sensitive
to X-rays in the range 15--1000 keV and has been in operation for nearly 12
years, it is a promising instrument to search for the cyclotron line.
We obtained an ISGRI spectrum  for each one of the SCW for which the source
appeared with a detection level above 7$\sigma$ and lay within 10 degrees
of the centre of the pointing. An average spectrum was created from the 37
resulting spectra.  The average spectrum obtained
from the 37 SCWs does not show any significant feature at the energy of the
previous reported line. A single power law component gives an acceptable
fit ($\chi^2=28$ for 21 degrees of freedom) in the range 20--180 keV. The
best-fit photon index is $2.65\pm0.08$.

Figure~\ref{pn-isgri} shows the combined EPIC-PN and ISGRI spectra. The
blackbody plus BMC and two narrow gaussians model gives a very good fit
(reduced $\chi^2\approx 1$). The best-fit values of the spectral parameters
obtained from the fit of this model to the joined EPIC-ISGRI spectrum are
consistent with those in Table~{\ref{specfit}, within errors. The energy
spectral index is slightly larger, 1.7, compared to 1.5 of the EPIC-PN
spectrum.

\section{Discussion}

\src\ is a persistent X-ray source with relatively low luminosity
($L_X \simless 10^{36}$) compared
to other accreting X-ray pulsars. It shows X-ray variability on timescales
of a few days with an amplitude of $\sim5$  \citep{reig09} and on
timescales of years by a factor of  $\simless 100$ 
\citep{ribo06}. The short-term variability has most likely an orbital
origin (a varying mass accretion rate due to its elliptical orbit,
$e=0.4$), while the long-term variability is probably due to oscillations
in the wind density from the donor star \citep{ribo06}.

Considerable progress has been made on the nature of this enigmatic system
in the past years. The optical companion is an O9.5Vp, where the {\em p}
indicates that it is a peculiar star. The peculiarity of \src\ is that  the
helium lines observed in the optical spectra \citep{blay06} and stellar
wind velocity \citep{ribo06} are stronger and slower, respectively, than
what is expected for a normal main-sequence O9.5 star. The highly debated
issue of the nature of the compact companion, whether it is a neutron star
or black hole, has been definitively solved in favour of the former with the
discovery of X-ray pulsations \citep{reig09}. 

\src\ has never been observed by any high-sensitivity instrument below 2
keV ({\it BeppoSAX} and {\it Suzaku} have $\sim$ 10 times less effective
area at $\sim0.3$keV and 1.5 keV, respectively than {\it XMM-Newton}). Thus
the {\it XMM-Newton} observations represent an excellent opportunity to
investigate the emission properties of \src\ in the soft X-ray band. 

%--------------------------------fig6----------------------------------------------
\begin{figure}
\resizebox{\hsize}{!}{\includegraphics{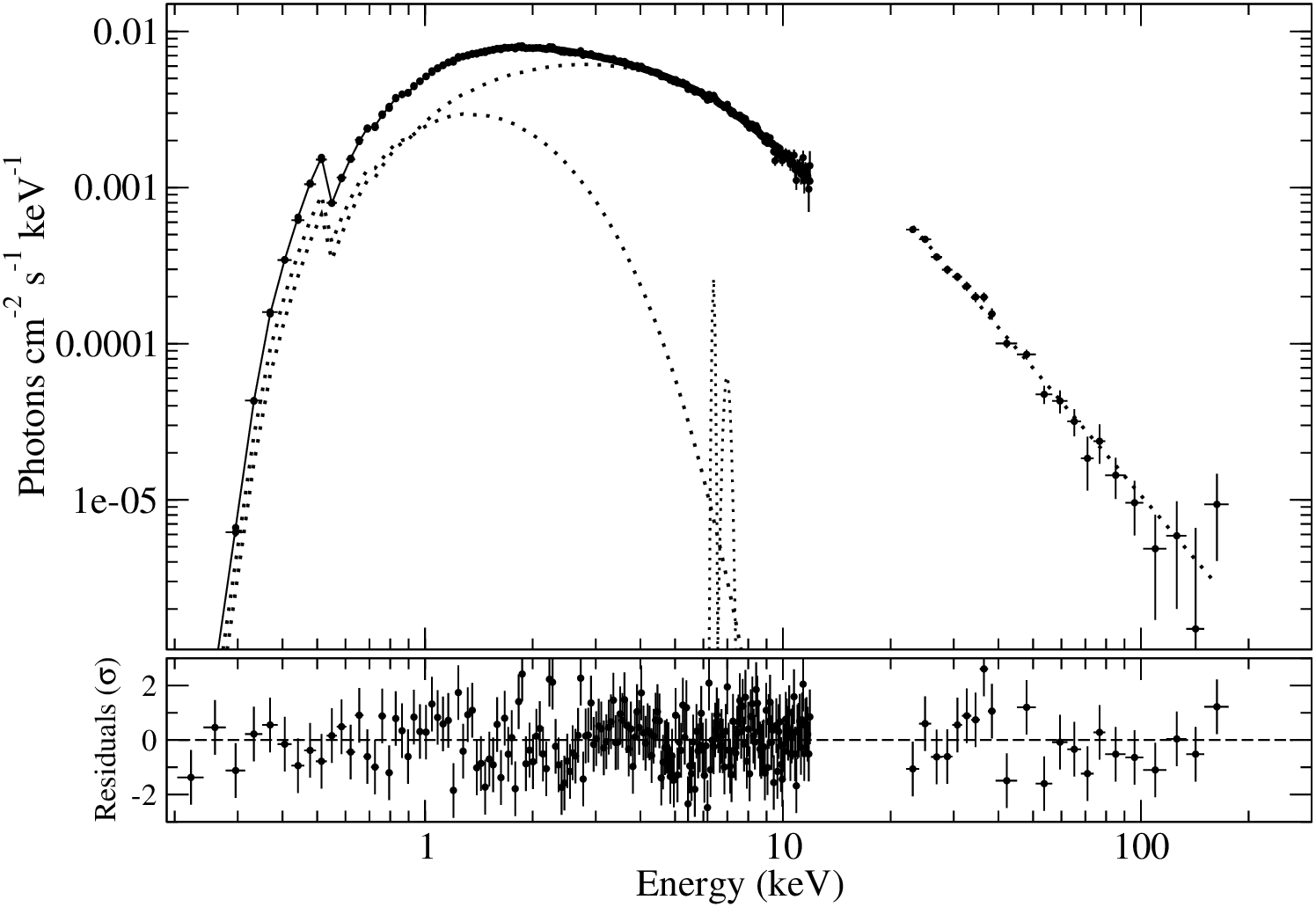} } 
\caption[]{Joined EPIC-PN and ISGRI spectra. The dotted lines correspond to
the blackbody plus bulk comptonization model of Table ~\ref{specfit}}
\label{pn-isgri}
\end{figure}
%------------------------------------------------------------------------------

\subsection{The source environment}

Fluorescent iron line emission  provides strong evidence for the presence
of material in the vicinity of the X-ray source. 
All previous observations of \src\ failed to detect any iron line at
or near 6.4 keV \citep{negueruela01,torrejon04,masetti04,reig09}. Upper
limits for narrow (0.1 keV) and broad (0.5 keV) lines had been set in $<56$
eV and $<65$ eV, respectively \citep{masetti04}. These results imply that 
there is not much material intrinsic to the system. The agreement between
the colour excess $E(B-V)$ derived from optical observations \citep{blay05}
and that obtained from our {\it XMM-Newton} observations (Sect.\ref{softx})
would support this result.

In a recent survey of HMXB with {\it Chandra}, \citep{torrejon10} showed that
the fluorescence Fe K$\alpha$ line is present in virtually 100\% of the
wind-fed HMXBs studied. The curve of growth shows that the reprocessing
material is a spherical shell of near neutral Fe, formed in the stellar
wind. It is surprising that although \src\ is a wind-fed system this line
has never been observed. The higher spectral resolution and
sensitivity of {\it XMM-Newton} below 10 keV compared to previous missions
allowed us to report the first detection of emission features at 6.4 keV
and 7.0 keV. These features can be interpreted as Fe K$\alpha$ and Fe XXVI
Ly$\alpha$ from photoionized plasma, respectively.

Although statistically significant, these lines are weak. Somehow the compact
object is not illuminating the stellar wind, as in the typical supergiant
systems, even though the accretion process is taking place, efficiently. 
The source shows random variations in the light curve typical of wind
accretors, as expected for a high-mass X-ray binary. Our results confirm
that the X ray source, however, is not deeply embedded into the stellar
wind of the mass donor, yet it is continuously accreting.  The large
orbital separation would imply that the wind is too tenuous when it reaches
the magnetosphere but the fraction of the stellar-wind mass-loss rate
captured by the neutron star would still be large enough to power the
X-rays.  This fraction is \citep[see e.g][]{bozzo08}

\[\dot{M}_{\rm capt}/\dot{M}_w \simeq R_{\rm acc}^2/(4 a^2) \]

\noindent where $a$ is the orbital separation and $R_{\rm acc}$ the
accretion radius, that is the distance at which the inflowing matter is
gravitationally focused toward the neutron star. It is given by

\[R_{\rm acc}= 2GM/v_w^2=3.7\times 10^{12}\, v_7^{-2} \,\,\, {\rm cm} \]

The stellar wind in \src\ is abnormally slow with $v=3.5\times 10^{7}$ cm
s$^{-1}$ \citep{ribo06}. Thus the accretion radius in \src\ is $R_{\rm
acc}= 3\times 10^{11}$ cm

The orbital period of \src\ is still not known. Two modulations in the
long-term X-ray light curve at 9.6 d and 19.2 d have been attributed to the
orbital period \citep{corbet07}. In either case, the neutron star is
orbiting at a very large distance from the donor  ($\sim 7R_{*}$ if $P_{\rm
orb}=9.7$ d and $\sim 10R_{*}$ if $P_{\rm orb}=19$ d). Assuming the radius of
\src\ to be $R_*=7.3 \rsun$ \citep{ribo06}, the orbital separation would
result in  $a=3.6\times 10^{12}$ cm  and $a=5.3\times 10^{12}$ cm for the
short and long orbital period solutions, respectively.

Thus the fraction of captured material would be $\dot{M}_{\rm
capt}/\dot{M}_w \simeq$ 0.002 and 0.001, respectively. The mass-loss rate
of an O9.5V star is $\sim 3 \times 10^{-8}$ $\msun$ yr$^{-1}$
\citep{ribo06}. Thus $\dot{M}_{\rm capt} \simeq 3 \times 10^{-11}$ $\msun$
yr$^{-1}$ or $2\times 10^{15}$ g s$^{-1}$, which is enough to power a
luminosity of $\sim L_X \approx GM_{\rm NS}\dot{M}_{\rm capt}/R_{\rm NS} \approx
3\times 10^{35}$ erg s$^{-1}$, as observed.

\subsection{The X-ray spectrum}
\label{xspec}

In order to investigate the nature of \src, it is illustrative to
compare its X-ray spectrum with that of other types of strongly magnetised
neutron-star systems: accreting X-ray pulsars and anomalous X-ray
pulsars (AXP). The former contain neutron stars with magnetic fields
$B\sim 10^{12}-10^{13}$ G, while AXPs are thought to harbour
magnetars with $B\sim 10^{13}-10^{15}$ G.

The spectra of many accreting X-ray pulsars show excess emission below
$\sim$2 keV. When this soft excess is modeled with a blackbody component,
the temperature and emission radius strongly depend on the source
luminosity \citep{hickox04}. In Be/X-ray binaries at low luminosity ($L_X
\simless 10^{35}$ erg s$^{-1}$), this soft component is believed to
represent emission from the polar caps. The main piece of evidence
supporting this interpretation is the size and temperature of the emitting
area. The radius of the emitting region is typically $\simless$ 0.5 km
\citep{mukherjee05,palombara06,palombara07,palombara09}, significantly
smaller than the assumed canonical size of the neutron star ($\sim10$ km).
Thus, isotropic emission from the surface of the neutron star is ruled out.
The blackbody temperature is $\simmore$ 1 keV. 

The X-ray spectra of AXPs in the 0.5--10 keV band are very soft and can
best be described by a blackbody ($kT\sim 0.5$ keV) plus a very steep
power-law model with $\Gamma \sim 3-4$ \citep{woods06,kaspi07} or by two 
blackbody components with characteristic temperatures $\simless 1$ keV
\citep{halpern05}.

The hard X-ray spectra (above 20 keV) of AXPs is characterised by a hard
power law tail extending up to 150 keV without any sign of turn over. The
power-law photon index range $\Gamma \sim 1-2$, that is, much harder than
the power laws seen below 10 keV \citep{gotz06}.  The X-ray spectrum of
accreting X-ray pulsars above 20 keV also exhibits power law tails.
However, these are significantly softer ($\Gamma \sim 2-3$) than those of
AXPs. Also, many accreting pulsars show exponential cutoffs at 30-50 keV
\citep{blay10}.

The spectrum of \src\ shows a soft excess at low energies. Irrespectively
of the model used, we find that the size of the emitting area in \src\ is
incompatible with the presence of an accretion disc and fits in the polar
cap scenario. The  temperature, $kT_{\rm bb}=1.63$ keV and radius, $R_{\rm
bb}=370$ m of the blackbody component (Table~\ref{specfit}) and the photon
index ($\Gamma=0.94$) are comparable to that seen in other low-luminosity
accreting pulsars \citep[][and references therein]{palombara09,reig09}.
However, unlike most accreting X-ray pulsars, but similarly to AXPs,
the hard power-law tail in
\src\ does not show any evidence for a cutoff below $\sim 150$ keV
(Fig.~\ref{pn-isgri}). The hard spectrum in \src\ is not as hard
as in AXPs, though.  

\subsection{Is \src\ a magnetar?}

\src\ has been observed by many space observatories. A cyclotron resonant
scattering feature (CRSF) at $\sim$30 keV has been reported in observations
by {\it RXTE} \citep{torrejon04}, {\it BeppoSAX} \citep{masetti04} and {\it
INTEGRAL} \citep{blay05}. All these detections were marginal (at $\simless
2\sigma$). \citet{wang09} analysed {\it INTEGRAL} data of \src\ and
reported a harmonic of this feature at $\sim$60 keV, although with very low
significance. The alleged CRSF was only detected during a high state
($L_X\approx 2\times 10^{35}$ erg s$^{-1}$) but not during a low state
($L_X\approx 4\times 10^{34}$ erg s$^{-1}$) in the data analysed by
\citet{wang09}. However, just a few days after the high state where the
detection of the harmonic was reported, \src\ exhibited two short-lived
hard flares with peak luminosity $L_X\approx 1.3\times 10^{36}$ erg
s$^{-1}$ and no CRSF was observed \citep{wang10}.  Quoted values of the
luminosity are for the 20--100 keV range and a distance of 2.6 kpc 
\citep{blay06}. Likewise, \citet{reig09} did not find any evidence for a
cyclotron line in the {\it RXTE} observations that led to the discovery of
the X-ray pulsations. The observations by \citet{reig09} are particularly
constraining because they analysed more than 138 ks worth of data with
instruments capable to detect a cyclotron line at 30 keV. The fact that
this feature, if real, is transient argues against its interpretation as a
CRSF.

On the other hand, the cyclotron line shape is a strong function of viewing
angle \citep{araya00}. Because of the long pulse period of \src, it might
be possible that the cyclotron line in \src\ is enhanced at certain pulse
phases and suppressed at others. When analysing a pulse-phase {\em
averaged} spectrum, the significance of the detection would depend on how
many spectra at the right pulse phase are used. We explored this idea by
performing pulse-phase {\em resolved} spectroscopy on the 2001 {\it RXTE}
observations analysed in \citet{torrejon04} and on ISGRI data taken on 
Rev. 383 (December 2005) and Rev. 510 (December 2006), but we did not find any significant change with pulse
phase. This kind of analysis is hampered by low statistics.

An independent estimate of the magnetic field strength can be obtained
considering the evolutionary track of the neutron star's spin period.
\citet{finger10} derived an average rate of the spin frequency change
between a {\it BeppoSAX} (November 1998) and a {\it Suzaku} (May 2007)
observation of  $\dot{\nu} = (-1.7\pm0.3) \times 10^{-14}$ Hz s$^{-1}$,
indicating that the rotation of the neutron star is slowing down. Based on
the slow pulsation period and the spin down rate, they show using a
magneto-rotation model for the evolution of neutron stars, that  the 5500 s
pulsations in the X-ray flux of \src\ can be explained provided the neutron
star in this system is a magnetar whose surface field at the present epoch
exceeds $10^{14}$ G. Indeed, the spin down rate is related to the magnetic
moment of the neutron star by \citep[see e.g][]{ikhsanov10}

\[\mu_m \approx 10^{32}k_t^{-1/2} (M_{1.5})^{1/2} I_{45}^{1/2}
\dot{\nu}_{-14}^{1/2} P_{5500} \, \, \, \, \,{\rm  G \,\,cm^3} \]

\noindent where where $I_{45}=I/10^{45}$ g cm$^2$ is the star's moment of
inertia, $k_t$ is a dimensionless parameter that accounts for the geometry
of the accretion flux and it is close to 1 for spherical accretion,
$\dot{\nu}_{-14}=\dot{\nu}/10^{-14}$ Hz s$^{-1}$ is the rate at which the
neutron star rotation changes with time, and $M$ is the mass of the neutron
star in units of 1.5$\msun$. From a detailed analysis of the 
magneto-rotational evolution of the neutron star in \src,
\citet{ikhsanov10} concluded that its surface magnetic field may range from
$5 \times 10^{13}$ G to   $3 \times 10^{15}$ G, depending on the geometry
of the accretion flux outside the magnetosphere, with a most probable value
in the interval $(6-9)\times 10^{13}$ G. This scenario also requires that
accretion in \src\ must occur without the formation of an accretion disc.
If an accretion disc were present then the observed evolution of the
rotation period in \src\ would imply a magnetic field strength larger than
$3\times 10^{15}$ G, which it is rather difficult to account for with
current models \citep{ikhsanov10}. The difficulty to accommodate an
accretion disc was already mentioned by \citet{torrejon04}. 

The  estimate of the magnetic field strength  based on the spin-down  rate
given by \citet{finger10} was determined using only two observations.  As
the authors state, "it is very unlikely that the spin-down rate obtained
from the {\it BeppoSAX} and {\it Suzaku} observations is valid on the long
term since the pulsar would reach zero frequency in 300 yr".  Surprisingly,
the spin-down rate obtained using all publicly available measurements of
the spin period and the new value from {\it XMM-Newton} agrees well with
that reported by \citet{finger10}. As can be seen in Fig~\ref{spinevol}, the
spin-down rate does not show evidence of a break-up. A linear fit to all
the measurements of the spin period of \src\ gives a  spin-down rate of
$\dot{\nu}=(-1.5\pm0.2) \times 10^{-14}$ Hz s$^{-1}$ with a correlation
coefficient of 0.97.  This rate implies a magnetic field of the order of
$B \sim 10^{14}$ G.

If the 1998 {\it BeppoSAX} data point, i.e., the oldest measurement, is
removed from the fit, then the spin-down rate decreases to half the
previous value, $\dot{\nu}=(-7.5\pm2.0) \times 10^{-15}$ Hz s$^{-1}$, which
would indicate that neutron star has started to rotate at a faster rate.
Still, the magnetic field derived from that spin-down rate (see equation
above) would be more than an order of magnitude stronger than common values
seen in accreting X-ray pulsars.

Recently, a new theory of quasi-spherical accretion in X-ray pulsars have
been developed \citep{shakura12}, where the dipole field in a wind-fed
neutron star is given by \citep[see e.g.][]{popov12}

\[B_{12} \sim 8.1 \dot{M}_{16}^{1/3} V_{300}^{-11/3} 
\left(\frac{P_{1000}}{P_{orb,300}} \right)^{11/12}\]

\noindent where $\dot{M}_{16}$ is the mass accretion rate in units of
$10^{16}$ gr s$^{-1}$, $V_{300}$ is the wind velocity in units of 300 km
s$^{-1}$, $P_{1000}$ is the spin period in units of 1000 seconds and
$P_{orb,300}$ is the orbital period in units of 300 days. Substituting the
values measured for 4U 2206+54: $\dot{M}_{16}=0.2$, $V_{300}=1.17$,
$P_{1000}=5.5$, and $P_{orb,300}=0.064$, the resulting magnetic field is $B
\sim 1.6\times 10^{14}$ G, in agreement with the magnetic field obtained
from the spin-down rate.

For a dipole magnetic field of $\sim 10^{14}$ G, the electron
cyclotron line would appear at $E > 500$ keV, where the photon statistics
are not good enough. On the other hand, a proton cyclotron line would
appear at $E \sim 0.5 (B/10^{14} G) = 0.5$ keV. Although a line with this
energy should be observable with {\it XMM-Newton} detectors, it is in a
region affected by strong interstellar absorption. Furthermore, no
significant lines have been detected in the persistent emission of
magnetars \citep{mereghetti08}.

The picture that emerges of \src\ from the available data at various bands
of the electromagnetic spectrum is that of an intriguing system: a highly
magnetized ($B\sim 10^{14}$ G) neutron star accreting matter from the
stellar wind of a O-type main-sequence star. Is \src\ a magnetar? \src\
does not fit in the traditional definition of magnetars, namely, {\em
very strongly magnetized ($B\sim 10^{14}-10^{15}$ G), isolated neutron
stars powered by magnetic energy}. First, the magnetic field in \src\ is in
the lower end and most likely below the canonical range of magnetic fields
in magnetars. Second, it is part of a binary system. Third, X-rays are
powered by accretion and the contribution to the X-ray luminosity from
dissipation of the magnetic field is expected to be two to three orders of
magnitude lower. 

The characteristics of \src\ would be closer to, but not fully
consistent with the definition of anomalous X-ray pulsars
\citep{mereghetti02}: {\em a spinning down pulsar, with a soft X-ray
spectrum, apparently not powered by accretion from a companion star, and
with a luminosity larger than the available rotational energy loss of a
neutron star}. As shown in Sect.~\ref{xspec}, the X-ray spectrum of \src\
differs from that of AXPs.

Alternatively, one can simply define a magnetar as a
neutron star with a magnetic field strength larger than the quantum
critical value $B_q=m^2cV3/\bar{h}e=4.4\times 10^{13}$ G at which the
energy between Landau levels of electrons equal their rest mass. In this
case, \src\ would contain a magnetar accreting matter from a luminous
companion. A more appropriate name for \src\ would then be {\em accreting
magnetar} \citep{ikhsanov10}.

\subsection{Accreting magnetars as a class of systems}

The idea
that wind-fed X-ray pulsars may contain super critical highly magnetised
neutron star has been discussed in previous works. The supergiant X-ray
binary 2S 0114+650 is the accreting X-ray pulsar with the longest period
(2.7 hr). No cyclotron line has been reported in this system. \citet{li99}
argue that, to interpret such a long spin period, the magnetic field
strength of this pulsar must have been initially $\simmore 10^{14}$ G.

\citet{bozzo08} showed that the large luminosity swings between quiescence
and outbursts (up to $\sim 10^5$) seen in supergiant fast X-ray transients
(SFXT) can be explained as the result of transitions across different
inhibition of accretion regimes. These regimes include inhibition due to
the magnetic and centrifugal barriers. In either case, magnetar-like fields
can be attained if the spin period of the system is long ($\simmore 1000$
s) or the system is luminous ($L_X\simmore 10^{36}$ erg s$^{-1}$).

Another candidate for accreting magnetar is the wind-fed supergiant X-ray
binary GX 301--2. \citet{doroshenko10} showed that a very strong magnetic
field ($2-3 \times 10^{14}$ G) is required to explain the long pulse
period of this system. A cyclotron line at 30--45 keV, implying a magnetic
field of the order $5 \times 10^{12}$ G \citep{kreykenbohm04,barbera05} has
been observed in GX 301--2. This magnetic field is well below the critical
value. However, these contradicting results can be reconciled if the
line-forming region is situated far above ($2-3 \, R_{\rm NS}$) the neutron
star's surface, i.e., in the accretion column \citep{doroshenko10}. Another
system that might belong to the class of highly-magnetised HMXB is IGR
J16358--4726, although its nature is much less certain \citep{patel07}

The main difference of \src\ with respect to 2S 0114+650, GX 301--2 and
SFXTs is that \src\ contains a main-sequence star, hence a still younger
system than those with supergiant companions. \src\ would be the natural
progenitor of supergiant X-ray binaries \citep{ribo06}. However, \src\ may
not be the only non-supergiant system with a likely highly-magnetised neutron
star. The magnetic field of the Be/X-ray binary X Per may also lie in the
magnetar range if a slow accretion wind is assumed \citep{doroshenko12}.

%%An important difference of 2S 0114+650 and the other systems is that its
%magnetic field was initially high, that is, it was born as a magnetar but
%then the magnetic field decayed to its present value  of $\sim10^{12}$ G. 
%GX 301--2, X Per and \src\ require a very strong field {\em at present}.

By comparing the common properties of these systems we can identify
the following general characteristics of accreting magnetar binaries:

\begin{itemize}

\item[--] They contain slowly rotating neutron stars  ($P_{\rm spin}\simmore
500$ s) with magnetic fields above the quantum critical value ($4.4 \times
10^{13}$ G) but lower than a few times $10^{14}$ G.

\item[--] They are persistent X-ray sources with luminosities in the range
$10^{35}-10^{36}$ erg s$^{-1}$ with long-term amplitude of X-ray
variability below two orders of magnitude.

\item[--] They are young systems ($\simless 10^6$ yr), undergoing spherical
wind accretion from luminous companions (OB stars) with abnormally low
stellar wind velocity ($\simless 400$ km s$^{-1}$).

\item[--] Although brief spin-up episodes may occur, spin down dominates over
the long-term with relatively short spin-down time scales $\tau\sim
P/\dot{P}\simless 10^3$ yr.

\end{itemize}

\section{Conclusion}

With the discovery of X-ray pulsations, the highly debated issue of the
nature of the compact companion has been solved, but have posed  new
questions. The combination of slow pulsations and long-term spin down of
the neutron star can be understood assuming that the magnetic field of the
neutron star is of the order of $10^{14}$ G. We have obtained a new value
of the spin-down rate and shown that it is consistent with  a supercritical
magnetic field. We searched for the presence of a cyclotron line in the
hard X-ray spectrum of \src, but we did not find any significant feature at
the previously reported energies.  \src\ adds up to a growing
population of highly-magnetised neutron stars in binary systems, with
magnetic field strengths in the magnetar range, whose X-ray emission is
powered by accretion. 

%If the high magnetic field of \src\ is finally confirmed, then \src\ would
%be the first accreting magnetar.  The soft X-ray spectrum is consistent
%with the presence of a soft excess component that is attributed to the
%emission from the polar caps. The temperature and size of the emitting
%region are comparable to that seen in other low-luminosity accreting
%pulsars, supporting the claim that this component may be ubiquitous in this
%type of systems.

\section*{Acknowledgments}

This work has been partially supported by the Spanish Ministerio de Ciencia
e Innovaci\'on through the projects AYA2010-15431, AIB2010DE-00054 (JMT) and 
20100026-ASIM
(PB). This work has made use of NASA's Astrophysics Data System
Bibliographic Services and of the SIMBAD database, operated at the CDS,
Strasbourg, France.

\label{lastpage}

\end{document}